\documentclass[times,11pt]{article}
\usepackage[french, english]{babel}
\usepackage{latexsym,amssymb,amsmath,amsfonts,amscd,color,epsfig,amsthm}

\usepackage[bold,full]{complexity}
\usepackage{pst-all}

\usepackage{float}
\floatstyle{ruled}
\newfloat{protocol}{tph}{lop}
\floatname{protocol}{Protocol} 
\newfloat{procedure}{tph}{lop}
\floatname{procedure}{Procedure} 
\usepackage[left=3.5cm,top=1cm,right=3.5cm,nohead,foot=1cm]{geometry}

\usepackage{tweaklist}
\renewcommand{\enumhook}{\setlength{\topsep}{0pt}  \setlength{\itemsep}{-1pt}}

\newcounter{def}
\newcounter{theo}
\newcounter{theo2}

\newtheorem{theorem}[theo]{Theorem}\def\TH{
\begin{theo}}\def\HT{
\end{theo}}
\newtheorem{theorem2}[theo2]{Theorem}\def\TH{
\begin{theo}}\def\HT{
\end{theo}}

\newtheorem{definition}[def]{Definition}\def\DE{
\begin{definition}}\def\ED{
\end{definition}}

\providecommand{\abs}[1]{\lvert#1\rvert}
\providecommand{\ceil}[1]{\lceil#1\rceil}

\begin{document}

\title{Exact, Efficient and Information-Theoretically Secure Voting
with an Arbitrary Number of Cheaters}

\author{Anne Broadbent$^{\,1,\,2}$\\
Stacey Jeffery$^{\,1,\,2}$\\
Alain Tapp$^{\,3}$\\
 \small 1. \sl Department of Combinatorics and Optimization, University of Waterloo, Canada\\
 \small 2. \sl Institute for Quantum Computing, University of Waterloo, Canada\\
 \small 3. \sl D\'epartment d'informatique et de recherche op\'erationnelle, Universit\'e de Montr\'eal, Canada
}
\date{15 November 2010}
\maketitle

\thispagestyle{empty}

\begin{abstract}

We present three voting protocols with unconditional privacy and
correctness, without assuming any bound on the
number of corrupt participants. All protocols have polynomial complexity
and require private channels and a simultaneous broadcast channel.
Unlike previously proposed protocols in this model,
the protocols that we present deterministically output the exact tally.
Our first protocol is a basic voting
scheme which allows voters to interact in order to compute the tally.
Privacy of the ballot is unconditional in the sense that regardless of the behavior of the dishonest participants nothing can be learned through the protocol that could not be learned in an ideal realisation. Unfortunately, a single dishonest participant can make the protocol abort, in which case the dishonest participants can nevertheless learn the outcome of the tally. Our second protocol introduces
voting authorities which improves the communication complexity by limiting interaction to be only
between voters and authorities and among the authorities themselves; the simultaneous broadcast is also limited to the authorities.
In the second protocol, as long as a single authority is honest,
the privacy is unconditional, however, a single corrupt authority or
a single corrupt voter can cause the protocol to abort.
Our final protocol provides a safeguard against corrupt voters by enabling a
verification technique to allow the authorities to revoke incorrect votes without aborting the protocol.
Finally, we discuss the implementation of a simultaneous broadcast
channel with the use of temporary computational assumptions,
yielding versions of our protocols that achieve everlasting security.

\vspace{.25cm}

\end{abstract}

%%%%%%%%%%%%%%%%%%%%%%%%%%%%%%%%%%%%%%%%%%%%%%%%%%%%%%%%%%%%%%%%%%%%%%%%%%%%%%%%%%%%%

\section{Introduction}

Multiparty secure computation enables
a group of~$n$ participants  to collaborate in order to compute a
function on their private inputs. Assuming that private random keys
are shared between each pair of participants,  every function can be
securely computed
if and only if less than~$n/3$
participants are corrupt; this fundamental result is due to  David
Chaum, Claude Cr\'epeau and Ivan Damg\a rard~\cite{CCD88} and to
Michael Ben-Or, Shafi Goldwasser and Avi Wigderson~\cite{BGW88}.
When a broadcast channel is available, the results of Tal Rabin and
Michael \mbox{Ben-Or}~\cite{RB89} tell us that this proportion can
be improved to~$n/2$.

Among all functions that can be computed with these general-purpose
protocols, perhaps the one that has the most obvious application is
voting. If we have a guarantee on the proportion of honest
participants, a secure voting protocol based only on pairwise
private channels can be implemented. 
(If, in addition to this, we have a broadcast channel, then we can
tolerate more cheaters). 
Here, we are interested in the case where no such guarantee on the proportion of honest participants is
available. Unlike some recent voting schemes  (see, for example, 
\cite{RS07,CGRV07,BHRSX09,CIVITAS,ScanII} or \cite{ARS06} for a review) our voting protocols are information-theoretically secure. In this sense, our schemes are not comparable to these more practical voting schemes. Implementation of our scheme in a practical voting situation is left for future work; in this paper we consider voting from a purely theoretical standpoint.

The first protocol for voting that is
information-theoretically secure even in the presence of a majority of
dishonest participants was presented in~\cite{BT07} and expanded in~\cite{WOTE}. Along
with the use of private communication, the protocol uses a
simultaneous broadcast channel.
In this paper, we present a new approach in the same model that achieves
better functionality.
Although our initial motivation was theoretical in nature, we
believe that this work may lead to interesting practical
applications.

We present three voting protocols. In our first protocol, we assume that each pair of voters is
connected by a private authenticated channel. In our second and third
protocols, we relax this assumption by  introducing \emph{voting
authorities}. The assumption then becomes that there are private and
authenticated channels only between voters and authorities and among
the authorities themselves. The third protocol improves on the second by enabling the authorities to revoke an invalid ballot.

All three protocols  require a  simultaneous broadcast
channel~\cite{CGMW85,HD05}, which, for our purpose, is  a collection
of broadcast channels where the input of one participant cannot
depend on the input of any other participant (see Definition~\ref{def:sim-broadcast}). This could be achieved
if all participants {\em simultaneously} performed a broadcast. In
the context of our second and third protocols, a simultaneous
broadcast among the authorities is sufficient.

\begin{definition}
\label{def:sim-broadcast}
An~$n$ participant simultaneous broadcast channel is a collection of~$n$ broadcast channels, one for each participant, such that
each participant chooses his input to the broadcast before receiving the value of any other participant's broadcast.
\end{definition}

It is not uncommon in multiparty computation to allow additional
resources, even if these resources cannot be implemented with the
threshold on the honest participants (the results of~\cite{RB89},
which combine a broadcast channel with~$n/2$ honest participants
being the most obvious example). Our work suggests that a
simultaneous broadcast channel is an interesting primitive to study
in this context. Given a resource to implement bit
commitment, we can implement a simultaneous broadcast: all
participants commit to their values, and then all participants open
these values. 
In Section~3, we discuss these issues further, and show that they lead to implementations of our protocols that provide \emph{everlasting} security. 

The last two of our protocols involve authorities that compute the tally and return its value to the voters. In order to do this, they can simply broadcast the tally to the voters. If the tally is not unanimous, the voters all learn that the protocol has failed.  Alternatively, a \emph{detectable Byzantine agreement} (broadcast) secure against faulty majorities \cite{FGHHS02}, which relies only on pairwise private channels can be used. The same type of channel is used for all broadcasts in all three protocols. 

Under the assumption that trapdoor one-way permutations exist, multiparty computation can be realized with no assumption on the number of dishonest participants \cite{FGHHS02}. The advantage of our scheme is that we only require simultaneous broadcast, which is no harder than bit commitment and can be implemented under the sole assumption of one-way functions.

\subsection{Goals and Assumptions}

Our voting protocols involve $n$~voters, each casting a ballot for a
single choice among $r$~candidates. The goal of the protocols is to
faithfully count the number of ballots in favour of each candidate
in such a way that voters' ballots remain private, honest ballots
are counted and dishonest voters cannot vote adaptively or influence the vote any more
than by honestly voting.
\textbf{Protocol~\ref{prot:vote-basic}} involves only the voters, while
\textbf{Protocols~\ref{prot:vote-authorities}} and~\textbf{\ref{prot:vote-authorities-robust}} also involve $t$ voting authorities.

We do not place any restriction on the dishonest participants, though we assume that all corrupt participants are part of a single collusion.
We present our protocols in the usual scenario where each voter
casts a ballot with a choice for a single candidate. Our protocols
can easily be adapted to allow any number of voices per ballot
(allowing, for instance, each voter to either choose two candidates,
or to vote twice for the same candidate). We can also add a dummy
candidate to allow voters to honestly cancel their ballots.

All three protocols are exclusively based on private authenticated
channels and a simultaneous broadcast channel. In the first
protocol, no assumption is made on the number of honest voters and
in the last two, the only assumption is that at least one authority
is honest. Under these assumption, our protocols provide perfect
privacy and correctness. This was believed to be
impossible~\cite{Jeroen} before~\cite{BT07} was published.
The major drawback is that any dishonest
participant can  make any protocol abort
(except in our third protocol, where only dishonest authorities
can make the protocol abort). When a protocol aborts, information about the tally may leak, but this never represents more information than would have been available, had the protocol succeeded.

\subsection{Comparison with Previous Work}
The first voting protocol that provides information-theoretic security in the presence of an unlimited number of cheaters was given in~\cite{BT07}. The protocol requires pairwise private channels and a simultaneous broadcast channel. It uses probabilistic techniques to evaluate the
tally for each candidate; for this reason, it is correct with
probability $1-2^{-\Omega(s)}$, with~$s$ being a chosen security parameter. Then, in~\cite{WOTE}, the protocol was extended to involve authorities. We use a similar technique to involve authorities in our work: the first idea is to use authorities to compute the tally, thereby restricting the number of participants in the simultaneous broadcast and reducing the number of required pairwise private channels. The second idea is to use the authorities to verify ballots, thereby preventing a voter from causing the protocol to abort.
This article improves on previous work in three ways. First, for honest participants, the outcome is deterministic and always represents the exact tally. Second, \textbf{Protocol \ref{prot:vote-basic}} is significantly more efficient than what was previously proposed. Finally, we believe that the idea of ``voting bins'' is more elegant and might have other applications.

\subsection{Summary of Results}
We now review the main features of each protocol. We omit here the statements ``except with exponentially small probability''. Formal protocols and their properties are given in Section~\ref{sec:vote}, and proofs of formal properties are given in Appendix~\ref{sec:proofs}.
%AT we an probabely not going to do that...
%AB we use the term "ideal" functionality below, so we need to say what it is.
It is common in multiparty computation to compare an implementation
of a functionality with its \emph{ideal} functionality. This ideal
functionality is represented as a black box, accepting private
inputs from each participant and privately communicating the
function evaluation on these private inputs back to each
participant.

\subsubsection{Basic Voting (Protocol~\ref{prot:vote-basic})}
\begin{itemize}
\item Only voters are involved in the protocol.
\item A collusion of dishonest voters can only learn through the protocol what they would learn in the
ideal functionality, and this even (and also) if the protocol aborts.
\item A single dishonest voter can make the protocol abort.
\item If all participants are honest, the protocol does not abort.
\item If the protocol does not abort, then the output is consistent with all ballots of the honest voters and
some assignment of ballots for the dishonest voters.

\item Dishonest voters cannot vote adaptively.
\end{itemize}

\subsubsection{Voting with Authorities (Protocol~\ref{prot:vote-authorities})}
\begin{itemize}
\item Voters and a small number of authorities are involved in the protocol.
\item Voters only interact with authorities and in a forward direction.
\item If at least one authority is honest, a collusion of dishonest voters and authorities
can only learn
what they would learn in the
ideal functionality, and this even (and also) if the protocol aborts.
\item A single dishonest voter or authority can make the protocol abort.
    \item If all participants are honest, the protocol does not abort.
\item If at least one authority is honest and if the protocol does not abort, then
the output is consistent with all ballots of the honest voters and some
assignment of ballots for the dishonest voters.
\item If at least one authority is honest, a collusion of dishonest voters and authorities cannot vote adaptively.
\end{itemize}

\subsubsection{Voting with Authorities and Verification (Protocol~\ref{prot:vote-authorities-robust})}
\begin{itemize}
\item Voters and a small number of authorities are involved in the protocol.
\item Voters only interact with authorities and in a forward direction.
\item If at least one authority is honest, a collusion of dishonest voters and authorities
can only learn
what they would learn in the ideal functionality, and this even (and
also) if the protocol aborts.
\item No collusion of voters alone can make the protocol abort.
\item A single dishonest authority can make the protocol abort.
\item If all participants are honest, the protocol does not abort.
\item If at least one authority is honest and if the protocol does not abort, then
the output is consistent with all ballots of the honest voters and some
assignment of ballots for the dishonest voters.
\item If at least one authority is honest, a collusion of dishonest voters and authorities cannot vote adaptively.
\item Dishonest voters not following the protocol will have their ballots revoked.
\item A dishonest authority can choose to revoke the ballot of an honest voter.
\item When a ballot is revoked, the voter who cast the ballot, as well as all authorities, know about it.
\end{itemize}

\subsection{Intuitive Description of the Protocols}
\label{sec:intuitive}

We now give a physical analogy to our protocols.
The protocols are modelled after the concrete setup of an array of~$rn$ bins, where~$r$ is the number of candidates, so that there are~$n$ bins per candidate. Each bin is
such that a voter may place a ball in any bin, but may not remove
a ball from a bin or observe the contents of a bin. Each voter is given a single ball to place in a single bin.
He randomly chooses one of his candidate's $n$ bins
and places his vote in said bin. When all votes have been cast, the
totals for each bin are revealed and each candidate's vote can be
tallied by summing over all $n$ of her bins. So far the need for $n$ bins per candidate is not clear, but we will soon see why it is necessary. 

For our protocols, we  model each bin as an integer \!\!$\pmod m$, with a vote consisting
of a string of~$rn$ integers, one integer for each bin. We choose $m= 2n+1$. The $i$th integer
of a vote represents the number of balls the voter places
in the $i$th bin. In this case, a well-constructed vote has
a single~$1$, and a value of~$0$ for all other bins. In our protocols,
each voter  splits his vote into shares, each share consisting
of~$rn$ integers~\!\!$\pmod m$, with the property that the bin-wise sum~\!\!$\pmod m$ of all the shares is equal to the vote. A vote that is split in this way is called a \emph{ballot}. Given a set of ballots shared among a group, it is easy to compute the tally without revealing any information on the individual votes: this is exactly what is required for a voting protocol!

Without looking at individual votes,  we must ensure that all votes that contribute to the tally are well-constructed. If a voter votes multiple times, there will be
extra votes, which will be detected in the tally stage. Our physical analogy breaks down at this point, since it is possible for a corrupt voter to cast a negative vote (by choosing  $m-1 \equiv-1 \pmod m$ votes) in some bin.  Thus, a cheating strategy would be to vote twice in one bin, and vote $-1$ in another.
However, having~$n$ bins makes it likely that most bins are empty,
and a negative vote in an empty bin causes it to have a negative
number of total votes. We define a negative number$\pmod m$ as a number whose residue$\pmod m$ is greater than $\frac{m}{2}$. Such
a number would be detected at the tally stage, since $\frac{m}{2}>n$,
and no bin can have more than $n$ votes if each voter votes once. This justifies the need for $n$ bins, as well as $m=2n+1$. A negative vote is detected with constant probability and repetition yields exponential security.

\subsection{Evaluation}

There are several properties by which a voting system may be evaluated
(see, for example, \cite{1029520}). In this section we give a high-level
evaluation of all three of our protocols with respect to these properties.

\textbf{Protocols~\ref{prot:vote-authorities}} and
\textbf{\ref{prot:vote-authorities-robust}}
make use of voting authorities. If we group the authorities
together, they act as a trusted third party, which means that
collectively they can violate privacy and correctness of the
protocol. However, taken individually, both privacy and correctness
are guaranteed as long as a \emph{single} authority is honest. This
suggests that in practice, authorities could be chosen to represent
different interest groups, with each voter needing to trust only a
single authority (note that it is not necessary for the voters to
trust the \emph{same} authority!). Strictly speaking,  \textbf{Protocol~\ref{prot:vote-basic}} does not involve authorities, however, when discussing general properties of all three protocols in this section, we consider the voters to be playing the role of the authorities in \textbf{Protocol~\ref{prot:vote-basic}}.

\subsubsection{Accuracy}

In all three protocols, as long as at least one authority is honest,
the tally is correct, or the protocol aborts. Also, if all participants are honest, the protocol succeeds and the tally is exact.

Each voter is guaranteed exactly one vote. The use of authenticated
channels permits each voter to cast a single ballot, and we can check
that each ballot contains only one vote by checking the total number
of votes in the tally (which should be equal to the number of ballots).
In \textbf{Protocol~\ref{prot:vote-authorities-robust}},
the verification step detects a ballot with more than one vote
with exponentially high probability, before the tally stage.

As long as a single authority is honest, no authority can add, change,
or delete a vote without the protocol aborting. In
\textbf{Protocol~\ref{prot:vote-authorities-robust}},
a single dishonest authority can cause a ballot to be revoked by
making it appear invalid. Unfortunately, the effect of this is extremely
similar to an authority being able to delete votes --- the only difference
being that all authorities and the voter are aware that
the vote is being deleted. Only the voter knows for certain that his vote was \emph{wrongfully} revoked.

Because of the ease with which an authority may cause a ballot to
be revoked, voter profiling could have a negative impact on the final
results of \textbf{Protocol~\ref{prot:vote-authorities-robust}}.
Although this is arguably the biggest flaw with \textbf{Protocol~\ref{prot:vote-authorities-robust}},
this is the tradeoff for ensuring that no voter can make the protocol
abort, and it may not be possible to circumvent this problem without
introducing new assumptions.

\subsubsection{Ballot Secrecy}

There are two aspects of ballot secrecy. The first is the privacy
of the vote. In all three protocols, if at least one authority is
honest, no information can be learned about a vote except what can
be learned from the tally.

The second aspect of ballot secrecy is a voting system's resistance
to voter coercion. If a voter can provide evidence to a third party
that he voted for a particular candidate, he can be coerced into voting
a certain way, or he can sell his vote. Unfortunately, in all three protocols,
a voter can sell his vote to an authority by voting 
in a specific
bin as instructed ahead of time by the buyer. Since most bins are
likely to be empty, the presence of a vote in a series of specified bins across all repetitions provides overwhelming evidence that the voter has complied.

\subsubsection{Verifiability}

In a general voting scheme, there are two things that a voter might
like to verify. The first thing is that his vote was properly cast, and the second is that his vote
was counted. In addition, the public may wish to verify that the system
is working correctly by performing an audit. In our protocols, if
there is a single honest authority, the protocol is certain to
be correct, or abort, and so there is nothing gained from verification,
since the protocol can't fail to be accurate if there is at least
one honest authority.

 The exception is in \textbf{Protocol~\ref{prot:vote-authorities-robust}}
where an authority can cause a ballot to be revoked. In this case, the authorities notify the voter via broadcast whether or not his vote was revoked.   This notification is guaranteed accurate provided at least one authority is honest.

If the voter is notified that his vote was not revoked, he can be sure that it is properly counted
or that the protocol aborts, as long as one authority is honest.
One way that the voter could verify that his vote was counted would
be if the tallies of each bin were made public. This is the case in \textbf{Protocol~\ref{prot:vote-basic}}, where the authorities
are just the voters. In this case, the voter checks to see that
there is at least one vote in the bin in which he voted and knows with
high probability that his vote was counted.

\subsubsection{Robustness}

It is often desirable for a voting system to be robust against breakdowns.
In all three of our protocols, a single dishonest authority can cause
the protocol to abort. In \textbf{Protocol~\ref{prot:vote-basic}}
and \textbf{Protocol~\ref{prot:vote-authorities}}, a single
dishonest voter can cause the protocol to abort, however this is improved
in \textbf{Protocol~\ref{prot:vote-authorities-robust}}
where no collusion of voters can make the protocol abort.

Although the protocols
are not  robust, their high accuracy and secrecy makes them useful
in situations where there is little to gain from making the protocol
abort. For example, in a federal election, the authorities would be chosen from respected members of various interest groups, so that each voter trusts at least one authority. In this case, although an authority might have much to gain from altering the vote, she would have nothing to gain from causing the protocol to abort. In fact, this would only serve to cast suspicion on all authorities, and perhaps a new group of authorities would be chosen.  The corrupt authority gains nothing from this outcome.

%%%%%%%%%%%%%%%%%%%%%%%%%%%%%%%%%%%%%%%%%%%%%%%%%%%%%%%%%%%%%%%%%%%%%%%%%%%%%%%%%%%%
\section{Voting Protocols}
\label{sec:vote}

We now present our protocols (Section~\ref{sec:protocols}) and formally state their properties (Section~\ref{sec:formal-theorems}). Proofs of the statements can be found in Appendix~\ref{sec:proofs}.

\subsection{Protocols}
\label{sec:protocols}

We now give our main protocols, \textbf{Protocols~\ref{prot:vote-basic}, \ref{prot:vote-authorities}}, and \textbf{\ref{prot:vote-authorities-robust}}, which make use of the various procedures  as given below. All protocols (and most procedures) may terminate with the outcome \emph{abort}; this is sometimes only implicit. If a called procedure aborts, then the calling procedure also aborts. This is also the case for the use of the simultaneous broadcast, which may abort if a participant refuses to participate. We say that a protocol \emph{succeeds} if it does not abort.

\textbf{Procedure~\ref{proc:random}} shows how  to generate randomness, in a group of participants~$S$. As long as one participant is honest, the output is an unbiased integer between~$0$ and $\ell-1$.

\begin{procedure} \caption{RANDOM} \label{proc:random}
{\bf Input:} $S,\ell$\\
{\bf Output:} $a \in \{0,...,\ell-1\}$
 \vspace{2pt}
 \hrule  \vspace{2pt}
\begin{enumerate}
\item Each participant~$i \in S$ sets $a_i \in_R \{0,...,\ell-1\}$\,.
\item Each participant~$i \in S$ inputs~$a_i$ into the simultaneous broadcast channel.
\item Each participant~$i \in S$ sets~$a=\sum_{i\in S} a_i \pmod \ell$.
\end{enumerate}
\end{procedure}

Our voting protocols (\textbf{Protocols~\ref{prot:vote-basic}, \ref{prot:vote-authorities}, \ref{prot:vote-authorities-robust}}) use a very basic $n$-out-of-$n$ threshold secret sharing scheme
called an additive secret sharing scheme.

\begin{definition}
An \emph{additive distributed  secret} $\text{ADS}(S,v,m)$ is a list of values $v_i \in \{0,\dots,m-1\}$,  each~$v_i$ in the possession of  participant~$i \in S$, such that \mbox{$v \equiv \sum_{i \in S} v_i \pmod m$} and such that any strict subset of the participants~$S$  has no information on~$v$. We call~$v$ the \emph{secret}. \end{definition}

A single participant can create an~$\text{ADS}(S,v,m)$ using \textbf{Procedure~\ref{proc:make-ass}}. It is easy to see that the procedure creates a valid ADS.

\begin{procedure} \caption{MAKE-ADS} \label{proc:make-ass}
{\bf Input:} $S,v,m$\\
{\bf Output:} $\text{ADS}(S,v,m)$
 \vspace{2pt}
 \hrule  \vspace{2pt}
\begin{enumerate}
\item For each $i \in \{1,...,\abs{S}-1\}$, set  $v_i \in_R \{0,...,m-1\}$. 
\item Set \mbox{$v_{\abs{S}} = v - \sum_{i=1}^{\abs{S}-1}v_i \pmod m$}. 
\item For all $i \in S$, privately send $v_i$ to participant~$i$\,.
\end{enumerate}
\end{procedure}

If participants share a set of~$k$~ADSs, they can, with no communication,  create a new ADS with secret being the sum \!\!$\pmod m$ of the initial ADSs. The method is given in \textbf{Procedure~\ref{proc:sum-ass}}. Note that this procedure does not allow the participants to learn any information about the contents of the ADSs.

\begin{procedure} \caption{SUM-ADS} \label{proc:sum-ass}
{\bf Input:} $\{X^j\}_{j=1}^k$ a set of ADSs, each of type $\text{ADS}(S, v_j, m)$\\
{\bf Output:} $Y$ an $\text{ADS}(S,v', m)$ where $v'= \sum_{j=1}^k{v_j} \pmod m$
 \vspace{2pt}
 \hrule  \vspace{2pt}
\begin{enumerate}
\item Each participant $i \in S$ sets $v_i' = \sum_{j=1}^k v_i^j \pmod m$.
\end{enumerate}
\end{procedure}

As mentioned in Section~\ref{sec:intuitive}, the main concept in our protocols is that  the voters distribute their votes in a system of bins: for each of the~$r$ candidates, there are~$n$ bins. Thus we call the list of all the bins for all candidates a VOTING-BIN-SET.

\begin{definition}
A VOTING-BIN-SET($S,n,r$) is a list of $rn$ \mbox{$\text{ADS}(S,v_{ij},2n+1)$} $(i= 1, \ldots , r, j = 1, \ldots , n)$.
\end{definition}

A BALLOT is a VOTING-BIN-SET that corresponds to a valid vote. This means that all ADSs have $v=0$ except a single one that has~$v=1$.

\begin{definition}
A BALLOT($S,n,r,c,o$) is a VOTING-BIN-SET($S,n,r$) such that only $v_{co}=1$ and all other $v_{ij}=0$.
\end{definition}

From above, we see that a BALLOT($S,n,r,c,o$) is a vote for candidate~$c$ expressed in bin number~$o$ of that candidate.
\textbf{Procedure~\ref{proc:sum-ballot}} is used to perform the bin-wise sum of the votes, such that the result is distributed as a VOTING-BIN-SET.

\begin{procedure} \caption{SUM-BALLOT} \label{proc:sum-ballot}
{\bf Input:} $\{X_j\}_{j=1}^n$ a set of BALLOTs\\
{\bf Output:} $Y$ a VOTING-BIN-SET
 \vspace{2pt}
 \hrule  \vspace{2pt}
\begin{enumerate}
\item Execute SUM-ADS for each ADS in each BALLOT to form the VOTING-BIN-SET~$Y$.
\end{enumerate}
\end{procedure}

\begin{definition}
A VOTING-BIN-SET is {\em sum consistent} if each ADS has secret $0 \leq v_{ij} \leq n$ and  \mbox{$\sum_{i \in [r], j \in [n]} v_{ij}  =n$}.
\end{definition}

Thus, as mentioned in Section~\ref{sec:intuitive}, a VOTING-BIN-SET that is \emph{not} sum consistent implies that the tally has been tampered with. We use this notion in our first voting protocol, \textbf{Protocol~\ref{prot:vote-basic}}. In all three protocols, $s$ is the security parameter.

\begin{protocol} \caption{VOTE-BASIC} \label{prot:vote-basic}
{\bf Input:} $\{x_i \}_{i=1}^n$, where $x_i \in \{0,\dots,r-1\}$ is the private vote for participant $i$\\
{\bf Output:} the tally for each candidate
 \vspace{2pt}
 \hrule  \vspace{2pt}
Let $S = \{0,\dots,n-1\}$\,. \\
Repeat the following in parallel $s$ times:
\begin{enumerate}

\item Each voter $i$ creates $X_i$, a BALLOT($S,n,r,x_i,o$)  where $o$ is randomly chosen in~$\{0,\dots,n-1\}$\,.
\item Voters create the sum, $Y=\text{SUM-BALLOT}(\{X_j\}_{j=1}^n)$\,.
\item Voters input their values for $Y$ into the simultaneous broadcast channel.
\item If $Y$ is not \emph{sum consistent}, abort.
\item The tally for each candidate is computed by adding all the values of the voting bins  in~$Y$ for the candidate.
\end{enumerate}
Each repetition should give the same answer, otherwise abort
\end{protocol}

In \textbf{Protocol~\ref{prot:vote-authorities}}, we use $S$ as a set of $t$ authorities and all ADSs, BALLOTs and VOTING-BIN-SETs
involve only the authorities.

\begin{protocol} \caption{VOTE-AUTHORITIES} \label{prot:vote-authorities}
{\bf Input:} $\{x_i \}_{i=1}^n$, where $x_i \in \{0,\dots,r-1\}$ is the private vote for participant $i$\\
\phantom{{\bf Input:}}   $S$, a set of authoritites\\
{\bf Output:} the tally for each candidate
 \vspace{2pt}
 \hrule  \vspace{2pt}
Repeat the following in parallel $s$ times:
\begin{enumerate}
\item Each voter $i$ creates $X_i$, a BALLOT($S,n,r,x_i,o$)  where $o$ is randomly chosen in~$\{0,\dots,n-1\}$\,.
\item Authorities create the sum, $Y=\text{SUM-BALLOT}(\{X_j\}_{j=1}^n)$\,.
\item Authorities input their values for $Y$ into the simultaneous broadcast channel.
\item If $Y$ is not \emph{sum consistent}, abort.
\item The tally for each candidate is computed by adding all the values of the voting bins in~$Y$ for the candidate.
\end{enumerate}
Each repetition should give the same answer, otherwise abort.\\
Authorities make the result of the vote public.
\end{protocol}

In \textbf{Protocol \ref{prot:vote-authorities-robust}}, we again use $S$ as a set of $t$ authorities and all ADSs, BALLOTs and VOTING-BIN-SETs involve only the authorities. This time, the authorities verify the BALLOTs. The voters will actually vote many times, each BALLOT being a valid vote to a random candidate.  Half of these ballots will be opened to test that they are valid. Then the voter will tell the authorities how to individually shift all the remaining BALLOTs so that they are all equal to his private vote. Protocol 5 is then used to test that all remaining BALLOTs vote for the same candidate, and one BALLOT is chosen. We repeat this $s$ times in parallel.

\begin{procedure} \caption{ADS-EQUALITY} \label{proc:ass-equality}
{\bf Input:} $\{X^j\}_{j=1}^{2s}$ a set of $2s$ ADSs of type $\text{ADS}(S,v_j, m)$\\
{\bf Output:} \emph{equal} or \emph{unequal}
 \vspace{2pt}
 \hrule  \vspace{2pt}
Repeat the following $s$ times in parallel:
\begin{enumerate}
\item The participants in $S$ use RANDOM to choose a random partition $\{P, Q\}$ of $\{X^j\}$ with $\abs{P} = \abs{Q} = s$\,.
\item The participants in $S$ use SUM-ADS to compute $Y=\sum_{i\in P} X^i - \sum_{i\in Q} X^i$\,.
\item The participants in $S$ input their values for $Y$ into the simultaneous broadcast channel.
\end{enumerate}
Return \emph{unequal} if any $Y$ has a secret that is not~$0$, otherwise return~\emph{equal}.
\end{procedure}

\begin{protocol} \caption{VOTE-AUTHORITIES-ROBUST} \label{prot:vote-authorities-robust}
{\bf Input:} $\{x_i \}_{i=1}^n$, where $x_i \in \{0,\dots,r-1\}$ is the private vote for participant $i$\\
\phantom{{\bf Input:}}   $S$, a set of authoritites\\
{\bf Output:} the tally for each candidate\\
\phantom{{\bf Output:}} for each voter, a bit indicating whether or not his vote was revoked
 \vspace{2pt}
 \hrule  \vspace{2pt}

For each voter~$i = 1, \ldots ,n$
\begin{enumerate}
   \item Voter $i$ creates $s$ \emph{sets} of $2s$ BALLOT($S,n,r,c,o$), where $c = x_i$ for each ballot and  $o$ is equal for all BALLOTs in a set, but random otherwise.
       Before distributing the BALLOTs, the voter \emph{encrypts} each BALLOT with two random \emph{shift} values: one between $0$ and $r-1$ that changes which candidate the BALLOT is for and the other between $0$ and $n-1$ that changes which bin the vote is in. The BALLOTs are then distributed.
   \item Each \emph{set} is treated separately at first. For each \emph{set}:
   \begin{enumerate}
      \item Half the BALLOTs are opened (chosen using RANDOM) and the voter's vote revoked if an opened BALLOT is not valid. Opened BALLOTS as discarded.
      \item For each remaining BALLOT, the voter broadcasts the \emph{shift} values. The authorities decrypt their shares of the remaining BALLOTs using these shift values.
   \end{enumerate}
   \item \label{step:ADS-EQUALITY} The authorities verify that each remaining BALLOT is a vote for the same candidate (without opening any vote). For each candidate $c$:
   \begin{enumerate}
      \item For each BALLOT, sum all the bins for candidate $c$ using SUM-ADS.
      \item Use $s$ rounds of ADS-EQUALITY to test equality of the above sums, the $i$th round having the $i$th and $i+1 \pmod s$th sets as input. Revoke the vote if any call to ADS-EQUALITY returns \emph{unequal}.
   \end{enumerate}
   \item The authorities broadcast to the voter and to the other authorities a single bit indicating whether or not his vote was revoked. If the  messages are not unanimous, abort.
\end{enumerate}
For each voter whose vote was not revoked, the authorities take a random remaining BALLOT from each \emph{set} (chosen using RANDOM) and use these to compute~$s$ parallel tallies as in \textbf{Protocol \ref{prot:vote-authorities}}.
Each repetition should give the same answer; otherwise abort.\\
Authorities make the result of the vote public.
\end{protocol}

\subsection{Formal Properties}
\label{sec:formal-theorems}

We now formally state the correctness, privacy and related  properties of our main protocols, \textbf{Protocols \ref{prot:vote-basic}}, \textbf{\ref{prot:vote-authorities}}, \textbf{\ref{prot:vote-authorities-robust}}, as well as our procedures.  In all theorems, we assume that if authorities are involved in the protocol, then
at least one authority is honest. Proofs of all theorems can be found in Appendix~\ref{sec:proofs}.

\begin{theorem}
ADS-EQUALITY detects inequality \!\!$\pmod m$ in $\{X^i\}$, except with exponentially small probability, as long as m is odd.
\end{theorem}

\begin{theorem}
\textbf{(Protocols \ref{prot:vote-basic}, \ref{prot:vote-authorities}, \ref{prot:vote-authorities-robust})} If all voters and authorities are honest, then the protocol succeeds and is correct with probability~1.
\end{theorem}

\begin{theorem}
\textbf{(Protocols \ref{prot:vote-basic}, \ref{prot:vote-authorities}, \ref{prot:vote-authorities-robust})} A collusion of dishonest participants  cannot learn more from the execution of the protocol than what they can learn from their inputs and the output of the ideal protocol.
\end{theorem}

\begin{theorem}
\textbf{(Protocols \ref{prot:vote-basic}, \ref{prot:vote-authorities}, \ref{prot:vote-authorities-robust})} Voters cannot vote adaptively.
\end{theorem}

\begin{theorem}
\label{thm:Correct-1-2}
\textbf{(Protocols \ref{prot:vote-basic} and  \ref{prot:vote-authorities})} Whatever the behaviour of a collusion of dishonest participants, the probability that the protocol succeeds and the output is inconsistent with the vote of the honest voters is exponentially small.
\end{theorem}

\begin{theorem}
\textbf{(Protocol~\ref{prot:vote-authorities-robust})}  Whatever the behavior of a collusion of dishonest participants, the probability that the protocol succeeds and the output is inconsistent with the vote of the honest voters is exponentially small.
Dishonest authorities can revoke votes.
\end{theorem}

In \textbf{Protocols \ref{prot:vote-basic}} and  \textbf{\ref{prot:vote-authorities}}, any participant can make the protocol abort, but this is not the case for  \textbf{Protocol~\ref{prot:vote-authorities-robust}}.

\begin{theorem}
\textbf{(Protocol~\ref{prot:vote-authorities-robust})} A voter can only make the protocol abort with exponentially small probability.
\end{theorem}

\subsection{Complexity}

All our protocols have polynomial computational and communication complexity. We now give the details of the communication complexity, which includes, for both the private and simultaneous broadcast channels, the number and size of messages sent per participant.

In \textbf{Protocol~\ref{prot:vote-basic}}, each voter's complexity is $n-1$ messages of size \mbox{$r n\lceil\log(2n+1)\rceil s$},
and one simultaneous broadcast of size $r n\lceil \log(2n+1)\rceil s$.
The protocol is done in two rounds of communication.

In \textbf{Protocol~\ref{prot:vote-authorities}}, each voter's complexity is $\abs{S}=t$ messages of size $r n\lceil\log(2n+1)\rceil s$, done in one round.
The authorities receive these messages and do one simultaneous broadcast of size $rn\lceil \log(2n+1)\rceil s$.
The protocol is done in two rounds of communication. In addition, making the result of the vote public requires for each authority a broadcast of size~$r \ceil{\log n}$\,.

In \textbf{Procedure \ref{proc:random}} (RANDOM), there is a single simultaneous broadcast of size $\lceil\log(\ell)\rceil$ among the $\abs{S}$ participants.

In \textbf{Procedure \ref{proc:ass-equality}} (ADS-EQUALITY), there is an initial call to RANDOM costing a simultaneous broadcast of size $s \lceil \log\frac{(2s)!}{s!s!}\rceil \in O(s^2\log s)$ among the participants in $S$, followed by another simultaneous broadcast of size $s \lceil \log m \rceil$. 

In \textbf{Protocol~\ref{prot:vote-authorities-robust}}, each voter's complexity is $t$ messages of size \mbox{$2rn\lceil\log(2n+1)\rceil s^2$}
followed by $t$ messages of size $s^2 \lceil\log(n r)\rceil$ done in two rounds.
The authorities receive these messages.  In opening half of the BALLOTs for each voter, they do a simultaneous broadcast of size $s^2 r n^2 \lceil\log(2 n + 1)\rceil$, and make one call to RANDOM (costing a single simultaneous broadcast of size $\lceil \log \binom{2s}{s}^{ns}\rceil \in O(n s^2 \log s)$).

In testing the equality of the remaining unopened BALLOTs, the authorities make $r n s$ calls to ADS-EQUALITY. Each of these calls costs two simultaneous broadcasts among the $t$ authorities, one of size $O(s^2\log s)$ and the other of size $s \lceil \log(2n+1) \rceil$.

In telling each voter whether or not his BALLOT was revoked, the authorities do $n$ broadcasts to $t+1$ participants, each of size~$1$.

The authorities make a single call to RANDOM (costing one simultaneous broadcast of size $\lceil\log s^{ns} \rceil = \lceil ns\log s \rceil$ among the $t$ authorities) to choose a BALLOT from each set for each voter.  They then do one more simultaneous broadcast of size $r n \lceil\log(2n+1)\rceil s$. In addition, making the result of the vote public requires for each authority a broadcast of size~$r \ceil{\log n}$\,.
 
In total, \textbf{Protocol \ref{prot:vote-authorities-robust}} requires: 
\begin{enumerate}
\item $tn$ secure authenticated channels, each of which will send $2$ messages of size $\leq 2rn\lceil \log(2n+1)\rceil s^2$ (since $r<n$ implies $s^2 \lceil \log(nr)\rceil < s^2 \lceil \log n^2 \rceil = s^2 \lceil 2\log n \rceil$).
\item $2rns+4$ simultaneous broadcasts among the $t$ authorities, each of size in $O(n^2 s^2 r \log (ns))$.
\item $n+1$ broadcasts from $t$ participants to $\leq n$ participants, each of size \mbox{$\leq r\lceil \log n \rceil$}. 
\end{enumerate}

%%%%%%%%%%%%%%%%%%%%%%%%%%%%%%%%%%%%%%%%%%%%%%%%%%%%%%%%%%%%%%%%%%%%%%%%%%%%%%%%%%%%%%
\section{Everlasting Security}
\label{sec:everlasting}

All three of our protocols rely on the existence of a simultaneous broadcast channel. In practice, such a channel can be obtained from a primitive called \emph{bit commitment}, \cite{53817} which  can itself be obtained either from one-way functions \cite{1250792,Nao91,HILL99} or based on the laws of physics~\cite{Kent99}.
Implementing a simultaneous broadcast channel using a bit commitment is simple: all participants commit to their values, and then all participants open these values.
Using this method with a computationally secure bit commitment yields \emph{everlasting
security}: as long as the computational assumptions are not broken during the execution of the protocol (more precisely, during the simulation of the simultaneous broadcast), the security of the protocols is perfect. Note that the privacy
of individual votes remains perfect even if these computational assumptions are broken during the protocol: breaking the simultaneous broadcast only helps voters or authorities to \emph{adaptively} modify the result of the vote.
In our context, because we cannot enforce  simultaneous commitment or simultaneous opening of the bit commitments, the use of an  unconditionally binding or an unconditionally concealing bit commitment is equivalent: breaking a computationally concealing bit commitment allows a participant to manipulate the tally  depending on the votes of those who have already committed to theirs, while breaking a computationally binding bit commitment allows a participant to manipulate the tally according to the commitments that occur afterwards.

%%%%%%%%%%%%%%%%%%%%%%%%%%%%%%%%%%%%%%%%%%%%%%%%%%%%%%%%%%%%%%%%%%%%%%%%%%%%%%%%%%%%%
\section{Future Work}
There are several areas of potential improvement to the protocols presented in this extended abstract. It is not known to us whether or not the protocols could be modified so that voter coercion is impossible. Another question of interest is whether or not it is possible to reduce the power of authorities to revoke honest votes without making additional assumptions. Also, we would not be surprised if the same functionality could be acheived more efficiently (under the same assumptions).

An interesting long term project will be to understand the functionalities that can be achieved under the simultaneous broadcast assumption (or even weaker assumptions). What
functionality beyond the ones presented here and those in \cite{BT07,WOTE} can be
achieved?  Simultaneous broadcast seems like a powerfull primitive and yet is achievable under physical assumptions \cite{Kent99}. Its power is not well understood  and we believe it warrants further exploration.

%%%%%%%%%%%%%%%%%%%%%%%%%%%%%%%%%%%%%%%%%%%%%%%%%%%%%%%%%%%%%%%%%%%%%%%%%%%%%%%%%%%%%

\bibliographystyle{alpha}
\bibliography{references}
\pagebreak
\begin{appendix}
\section{Proofs of Formal Properties}
\label{sec:proofs}

\begin{theorem2}
ADS-EQUALITY detects inequality \!\!$\pmod m$ in $\{X^i\}$, except with exponentially small probability, as long as m is odd.
\end{theorem2}
\begin{proof}

Suppose the input $\{X^j\}$ is unequal. Let $P$ and $Q$ be any partition of $\{X^j\}$ such that $\abs{P}=\abs{Q}$ and $\sum_{i\in P} X^i = \sum_{i\in Q} X^i$.  Note that by swapping any two non-equal elements in $Q$ and $P$ respectively, we make the two sums unequal.  This observation is not entirely obvious, since we're working \!\!$\pmod m$.  Suppose we swap $a \in P$ and $b \in Q$ where $a > b$.  This will result in $\sum_{i \in P} X^i$ decreasing by $a-b$ and $\sum_{i \in Q} X^i$ increasing by $a-b$. Since these two sums were equal before the swap, we now have a difference of $2(a-b)$. If $2(a-b) \equiv 0 \pmod m$, since $m$ is odd, then we must have $(a-b) \equiv 0 \pmod m$, which is a contradiction since $a$ and $b$ are assumed non-equal.  So as long as we swap non-equal elements from the partitions, their sums will no longer be equal.  From this observation, we will show that there are at least as many partitions with $\sum_{i\in P} X^i \neq \sum_{i\in Q} X^i$ as there are with $\sum_{i\in P} X^i = \sum_{i\in Q} X^i$.

Consider the operation of swapping the first two unequal elements in the sorted sets $P$ and $Q$. Clearly this operation maps equal sum partitions to unequal sum partitions. In addition, let us specify that the partition where $P$ and $Q$ are identical be mapped to the partition obtained by sorting the set $\{X^i\}$ and setting $P$ equal to the first half; the result will be unequal since $\{X^i\}$ is unequal.   We now have a one-to-one mapping from equal sum partitions to unequal partitions, so no more than half of the possible partitions can have the property $\sum_{i\in P} X^i = \sum_{i\in Q} X^i$.

Thus the probability of choosing a partition with this property when two or more elements are unequal is less than $\frac{1}{2}$. With $s$ repetitions, the probability of an unequal set passing ADS-EQUALITY is less than $\frac{1}{2^{s}}$.
\end{proof}

\begin{theorem2}
\textbf{(Protocols \ref{prot:vote-basic}, \ref{prot:vote-authorities}, \ref{prot:vote-authorities-robust})} If all voters and authorities are honest, then the protocol succeeds and is correct with probability~1.
\end{theorem2}
\begin{proof}
This follows from the additivity of the secret sharing scheme.
\end{proof}

\begin{theorem2}
\textbf{(Protocols \ref{prot:vote-basic}, \ref{prot:vote-authorities}, \ref{prot:vote-authorities-robust})} A collusion of dishonest participants  cannot learn more from the execution of the protocol than what they can learn from their inputs and the output of the ideal protocol.
\end{theorem2}
\begin{proof}
For \textbf{Protocol \ref{prot:vote-basic}} and \textbf{\ref{prot:vote-authorities}}, since we assume that there is at least one honest authority (or one honest voter in the case of \textbf{Protocol~\ref{prot:vote-basic}}), this follows from the properties of the secret sharing scheme, and the fact that SUM-ADS reveals no information about the values of the ADSs. We give a slightly more in depth discussion for \textbf{Protocol \ref{prot:vote-authorities-robust}}.

Upon receiving the distributed BALLOTs, the authorities open half of the BALLOTs in each set. Since these BALLOTs are encrypted using random shifts, any opened BALLOT votes for each of the candidates with probability $\frac{1}{r}$, independent of the voter's chosen candidate, $x_i$. The shift values are only revealed for the unopened BALLOTs, and the application of shifts to a BALLOT by each authority simply results in the authorities having the correct BALLOT as an ADS. We know that as long as one authority is honest, no information can be learned about the BALLOT by any authority.

Next the authorities use several rounds of SUM-ADS and ADS-EQUALITY to verify the equality of the unopened votes. We noted earlier that SUM-ADS does not allow participants to learn anything about the contents of the ADSs. As for ADS-EQUALITY, in each of $s$ rounds, the value of \mbox{$Y=\sum_{i\in P}X^i - \sum_{i\in Q}X^i$} is revealed. If a voter is honest, all of his values for $Y$ will be $0$, and so no information will be revealed, unless an authority has altered his share of some of the ADSs. Say an authority (or similarly, a group of authorities) changes his (their) shares of the $\{X^j\}_{j=1}^{2s}$ by adding $a_j$ to his share of $X^j$ for $j=1\dots 2s$. Then, assuming the $\{X^j\}_{j=1}^{2s}$ were equal before, we now get \mbox{$Y=\sum_{i\in P}a_i - \sum_{i\in Q}a_i$}. This value reveals nothing about the value of any $X^j$. 
\end{proof}

\begin{theorem2}
\textbf{(Protocols \ref{prot:vote-basic}, \ref{prot:vote-authorities}, \ref{prot:vote-authorities-robust})} Voters cannot vote adaptively.
\end{theorem2}
\begin{proof}
This follows from the use of simultaneous broadcast.
\end{proof}

\begin{theorem2}
\label{thm:Correct-1-2}
\textbf{(Protocols \ref{prot:vote-basic} and  \ref{prot:vote-authorities})} Whatever the behaviour of a collusion of dishonest participants, the probability that the protocol succeeds and the output is inconsistent with the vote of the honest voters is exponentially small.
\end{theorem2}
\begin{proof}
Suppose a collusion of $k<n$ dishonest voters wishes to cause the final tally to be incorrect. If they vote more than $k$ times between them, the final tally will be greater than $n$ and the protocol will abort. Thus, the only way for them to cause the final tally to be incorrect is for at least one of them to vote negatively in at least one bin, say bin $b$. If no other voter casts a vote in bin $b$, then the bin total will be $m-1 > n$, so the protocol will abort. For a voter to succeed in voting negatively without the protocol aborting, he must vote negatively in a bin with at least one vote in it.

Even in the worst case, where all $n-1$ other voters vote for the candidate to which bin $b$ belongs, the probability that bin $b$ is empty is $(\frac{n-1}{n})^{n-1}$, which is greater than $\frac{1}{3}$ for $n \geq 2$. By repeating the protocol $s$ times, the probability that a voter successfully casts a single negative vote without the protocol aborting is less than $(\frac{2}{3})^{s}$.

A similar analysis applies to corrupt authorities in \textbf{Protocol~\ref{prot:vote-authorities}}: since there is at least one honest authority, any collusion (even involving authorities) that modifies or casts an incorrect vote will cause the protocol to abort, except with exponentially small probability. 
\end{proof}

\begin{theorem2}
\textbf{(Protocol~\ref{prot:vote-authorities-robust})}  Whatever the behavior of a collusion of dishonest participants, the probability that the protocol succeeds and the output is inconsistent with the vote of the honest voters is exponentially small.
Dishonest authorities can revoke votes.
\end{theorem2}
\begin{proof}
The proof that a collusion of dishonest voters cannot cause the output to be inconsistent with the vote of the honest voters (with exponential probability) is the same as that given in \textbf{Theorem \ref{thm:Correct-1-2}}, since in \textbf{Protocol \ref{prot:vote-authorities-robust}}, voters have even \emph{less} power to cheat, due to the verification steps which distinguish \textbf{Protocol \ref{prot:vote-authorities-robust}} from \textbf{Protocol \ref{prot:vote-authorities}}. If the authorities attempt to modify a BALLOT, just as in \textbf{Protocol \ref{prot:vote-authorities}}, they must subtract a vote from some bin for every vote they add to any bin. This has an exponentially high probability of causing a negative total in some bin just as it does in the case of \textbf{Protocol \ref{prot:vote-authorities}}, since the distribution of votes in the bins for a particular candidate are independent from set to set.

The only extra power of the authorities to affect the outcome through the added verification is their ability to revoke votes. A dishonest authority can cause an honest vote to be revoked by simply adding $1$ to his share of any bin of a BALLOT selected for opening. This will cause the BALLOT to appear invalid, and the vote will be revoked. 
\end{proof}

\begin{theorem2}
\textbf{(Protocol~\ref{prot:vote-authorities-robust})} A voter can only make the protocol abort with exponentially small probability.
\end{theorem2}
\begin{proof}
For a voter to make the protocol abort, he must cast a vote that is not a valid BALLOT, or cast votes across the $s$ parallel tallies for different candidates.

If a voter wants to cast an invalid BALLOT in a set, he must create no more than $s$ invalid BALLOTS in that set, or he is guaranteed to have at least one invalid BALLOT opened at step (3) of the protocol. If he has $1 \leq x \leq s$ invalid BALLOTs, then the probability that no invalid BALLOT is opened is:
$$
\frac{\binom{2s-x}{s}}{\binom{2s}{s}} = \frac{s(s-1) \dots (s-x+1)}{2s(2s-1) \dots (2s-x+1)} = \frac{1}{2} \prod_{i=1}^{x-1} \frac{s-i}{2s-i} \leq \frac{1}{2}\,.
$$
If the voter has at least one invalid BALLOT per set, the probability that an invalid BALLOT is not opened is $\leq \frac{1}{2^s}$.
If the voter has invalid votes in some sets and not others, then his BALLOTs are not equal. Here we take a looser definition of equal, whereby two BALLOTs are equal if and only if each BALLOT has the same number of votes for each candidate. This is the type of equality required so that the final tallies of each round are equal. If the unopened BALLOTs are not equal in this sense, then with exponentially high probability, ADS-EQUALITY will return \emph{unequal} at step~\ref{step:ADS-EQUALITY} and the vote will be revoked. Hence, except with exponentially small probability, the voter cannot make the protocol abort.
\end{proof}

\end{appendix}
\end{document}